\documentclass[a4paper]{spie}  
\usepackage[]{graphicx}

\title{Ratchets in homogeneous extended systems: internal modes and the
role of noise}

\author{Angel S\'anchez\supit{a}, Luis Morales-Molina\supit{b}, Franz G.\ Mertens\supit{b},
Niurka R.\ Quintero\supit{c,d},
Javier Buceta\supit{e} and Katja Lindenberg\supit{e}
\skiplinehalf
\supit{a}Grupo Interdisciplinar de Sistemas Complejos (GISC) and Departamento 
de Matem\'aticas, Universidad Carlos III de Madrid, 28911 Legan\'es, Madrid, Spain\\
\supit{b}Physikalisches Institut, Universit\"at Bayreuth, D-85440 Bayreuth, Germany\\
\supit{c}Departamento de F\'\i sica Aplicada I, E.\ U.\ P.,
Universidad de Sevilla, Virgen de \'Africa 7, 41011 Sevilla, Spain\\
\supit{d}Instituto Carlos I de F\'{\i}sica Te\'orica y
Computacional, Universidad de Granada,\\ 18071 Granada, Spain\\
\supit{e}Department of Chemistry and Biochemistry, and Institute for Nonlinear
Science,\\ University of California, San Diego, 9500 Gilman Drive, La Jolla,
California 92093-0340, U.S.A.  
}

\authorinfo{Further author information: (Send correspondence to A.S.)\\A.S.: E-mail:
anxo@math.uc3m.es. URL: http://gisc.uc3m.es/$\sim$anxo\\L.M.-M.: E-mail: 
luis.morales-molina@uni-bayreuth.de\\F.G.M.: E-mail: franz.mertens@uni-bayreuth.de
\\N.R.Q.: E-mail: niurka@euler.us.es
\\J.B.: E-mail: jbuceta@hypatia.ucsd.edu\\K.L.: E-mail: kl@hypatia.ucsd.edu}

\begin{document} 
\maketitle 

\begin{abstract}
We revisit the issue of 
directed motion induced by zero average forces in extended systems
driven by ac forces. It has been shown recently that a
directed energy current appears if the ac external force, $f(t)$, breaks the
symmetry $f(t) = - f(t+T/2)$, $T$ being the period, if topological solitons
(kinks) existed in the system.
In this work, a collective coordinate approach allows us to identify the
mechanism through which the width oscillation drives the kink and its relation
with the mathematical symmetry conditions. Furthermore, our theory predicts,
and numerical simulations confirm, that the direction of motion depends on
the initial phase of the driving, while the system behaves in a ratchet-like
fashion if averaging over initial conditions. Finally, the presence of
noise overimposed to the ac driving does not destroy the directed motion;
on the contrary, it gives rise to an activation process that increases the
velocity of the motion. We conjecture that this could be a signature of 
resonant phenomena at larger noises.
\end{abstract}



\section{INTRODUCTION}

Within the exciting realm of transport phenomena in nonlinear systems
\cite{Scott}, net directed motion induced by zero average forces is
a very intriguing phenomenon that is being the subject of much research 
in the last ten years (see Ref.\ \citenum{Reimann} for a review; 
see also Ref.\ \citenum{Hanggi_a} for an introductory presentation).
The interest on this issue stems from the attempts to understand the 
way transport takes place inside cells through the so-called 
molecular motors \cite{Prost}. Also, nanotechnology has motivated a 
number of works on this kind of phenomena, in view of its possible 
applications as current rectifiers \cite{Linke}. Although the first 
works on this topic can be dated back to the beginning of the 
20th century (see Sec.\ 1.2 in Ref.\ \citenum{Reimann} for an interesting historical
recollection), it was Feynman\cite{Feynman} who made the seminal 
contribution, after which the name {\em ratchet} is generally used to 
refer to this very general problem.

At a first stage, the ratchet phenomenon was studied in a zero-dimensional
context, following the schemes proposed in Refs.\ \citenum{Magnasco,Prost2}, 
in which a point particle was subjected to the action of a periodic, 
asymmetric potential and a periodic force\cite{Magnasco} or, alternatively,
the potential was periodically switched on and off \cite{Prost2}. Subsequently,
the statistical physics community devoted a great deal of effort to the 
subject, and very many types of ratchet were proposed and studied (see 
Refs.\ \citenum{0d} for a very short list of examples; see \citenum{Reimann}
for a very exhaustive bibliography).

The next step in the search for directed motion scenarios was to consider
extended systems, focusing specially in soliton-bearing models (or, more
generally, systems where nonlinear coherent excitations play a key role 
for the transport properties). Indeed, the fact that in many situations 
solitons behave very much like particles \cite{Scott,SIAM} led immediately
to search for generalizations of the one-degree-of-freedom results
to more (even infinite)
degrees of freedom. Among the first
works in this direction, we could mention Refs.\ 
\citenum{Marchesoni,Tsironis,salquin}, which studied solitons in a noisy environment.
Again, much as in the case of point-like ratchets, these proposals stimulated
both the statistical mechanics and the nonlinear science communities to 
look for further examples, not only theoretical 
\cite{ext-th},
but also experimental and applied \cite{ext-app}. 

\section{Homogeneous extended ratchets}

Generally speaking, the appearance of ratchet-like behavior requires
two ingredients: departure from thermal equilibrium (either by using
correlated stochastic forces or deterministic forces) and breaking of
spatial inversion symmetry \cite{Reimann}, as mentioned above. This 
is actually the setup in the systems studied in the references cited
so far and in a majority of other ratchet models. However, it has 
recently been realized that the use of an {\em asymmetric driving}
can play the same role as the spatial asymmetry. Such an effect was 
first 
proposed for one-particle
systems in Ref.\ \citenum{Flach2} (see also Ref.\ \citenum{Hanggi_b}). The 
analysis presented by Flach and coworkers indicated that a
directed energy current appeared
if $f(t)$ broke the symmetry $f(t) = - f(t+T/2)$, $T$ being the period
of the external driving. It was only natural then to try to verify this
phenomenon in extended systems, and so it was done both in the quantum 
\cite{Hanggi_c} and in the classical cases \cite{Flach,SZ}. In this work,
we revisit the example studied by these last authors, namely the ac 
driven, damped, sine-Gordon equation, given by

\begin{equation}
\phi_{tt} - \phi_{xx} + \sin(\phi) = -\beta \phi_{t} + f(t).
\label{sG}
\end{equation}

In Refs.\ \citenum{Flach,SZ}, the previous symmetry considerations were
generalized to the extended model (\ref{1}). Again, it was found that
if $f(t)$ broke the symmetry $f(t) = - f(t+T/2)$, {\em and the total 
topological charge in the system was nonzero}, a directed current should
be observed whose direction and magnitude depended on the driving and
damping parameters. It is important to realize that the condition on 
the topological charge implies that {\em at least one kink or one
antikink must exist}, and that the numbers of both types of nonlinear 
excitations should differ at least by 1. In the case when there is just
one kink (or antikink) in the system, we are faced with an analogy with 
the point-like ratchet proposed in Ref.\ \citenum{Flach2}. Indeed, as in many
other instances \cite{SIAM}, kinks behave basically as point-like particles,
and the fact that their presence in the extended system is needed to 
have directed current reinforces this analogy. However, as we will see 
immediately, the scenario is not that simple, and in fact the point-like
particle picture is not enough to understand the phenomenology. 

As their working example, Flach {\em et al.}\cite{Flach}
and Salerno and Zolotaryuk\cite{SZ} considered 
$f(t)\equiv \epsilon_{1} \cos(\delta t) + \epsilon_{2}
\cos(m \delta t + \theta)$. For this choice, they performed numerical 
simulations that confirmed the symmetry analysis results. In view that
the system did exhibit ratchet-like behavior, i.e., it rectified 
ac current, as kinks moved towards one direction in space, they tried
a collective coordinate approach (see Ref.\ \citenum{SIAM} for a 
review on this technique), in which the kink motion was reduced to a
description in terms of an ordinary differential equation for the
motion of its center. However, their attempt turned out to be 
unsuccesful \cite{Flach,SZ},
showing that, as advanced above, there was more to the
net directed motion observed than the analogy to the point-like 
ratchet. 

\section{What's wrong with this particle?}

At this point, it is interesting to discuss in some detail the 
context in which this research should be understood from the 
viewpoint of kink dynamics. A few years earlier, some of us 
\cite{Niurka1} have proven that sine-Gordon kinks can not move 
when subjected to single-harmonic, ac forces plus damping. This 
had been debated during the nineties but the issue was finally 
settle down in the negative sense. For that problem, a collective 
coordinate description in terms of the motion of the kink center
showed a quantitative agreement with the numerical simulation
results. In the absence of damping, net directed motion was 
possible with single-harmonic forces, and again the phenomenon was
very accurately captured by the collective coordinate approach. 

In a subsequent work, we applied the same approach to a related system,
the so-called $\phi^4$ equation \cite{Niurka2,Niurka4}. In this case,
we found that, in spite of the fact that for most parameter choices 
the kink behavior was correctly described by a particle picture, there
were anomalous resonances that could only be described by taking into
account that $\phi^4$ kinks possess an internal mode, namely an 
oscillation of the kink shape arising from simple linear stability 
analysis. To include such an internal mode in the theory, we resorted
to an improved collective coordinate approach with two variables, 
namely the position and the width of the kink, thus recovering the
complete agreement with the simulations. 

In view of this, it seems natural to think that something similar 
was taking place when the sine-Gordon system was forced with two 
harmonics. In the original works \cite{Flach,SZ}, the authors proposed
a similar explanation for the failure of their approach.
An early 
attempt to develop a two-variable theory for this problem\cite{Willis}
failed as well, as it could only provide a partial answer indicating 
that it was indeed necessary that the kink width was a dynamical 
variable. Furthermore, the correct description of the dynamics and the 
connection to the symmetry analysis were still lacking. But there was
yet another problem in this line of reasoning: that sine-Gordon kinks
do not exhibit internal modes, as was shown in Ref.\ \citenum{Niurka3}.
Nevertheless, such a conundrum might be solved if we realized that, in
fact, the sine-Gordon equation we are studying is forced, and it has
been shown \cite{Niurka5} that external forces can excite phonons 
(linear radiation waves) that give rise to a total effect similar to
that of an internal mode. 
Thus, it was later shown
in the case of the soliton ratchet proposed by Salerno and 
Quintero (an inhomogeneous sine-Gordon equation, see Ref.\
\citenum{salquin}) that the crucial factor was the interaction
of that effective
internal mode and the translation mode. Another instance where the
possible role of internal modes had been previously highlighted was Ref.\ 
\citenum{Cilla}, where a two-particle system with an internal degree of 
freedom was studied, exhibiting ratchet behavior in symmetric potentials. 
In view of all this,
the possibility of trying to incorporate an internal mode in the description
of this problem seemed to be an intriguing issue. This we discuss in 
the next section. Some of the results below were presented in a much 
more concise form in Ref.\ \citenum{prl}.

\section{Including the internal mode}

In the following, we consider again the system proposed in Refs.\ 
\citenum{Flach,SZ}. In order to generalize their work, we consider a more
general form for the ac driving, namely
\begin{equation}
\label{fuerza}
f(t)\equiv \epsilon_{1} \sin(\delta t + \delta_{0}) + \epsilon_{2}
\sin(m \delta t + \delta_{0} + \theta),
\end{equation}
noting that the original formulation is recovered by setting $\delta_0=
\pi/2$.
Hereafter, we will refer to $\delta_0$ as the {\em initial phase} and
to $\theta$ as the {\em relative phase}.

Our collective coordinate theory is based on
an {\em Ansatz}, proposed in Ref.\ \citenum{Rice},
for the perturbed kink depending on two variables, $X(t)$
and $l(t)$ (position and width of
the kink). It is not difficult to show \cite{Rice,Niurka2,Niurka3}
that the dynamics of these two collective coordinates is given by
\begin{eqnarray} \label{cc-p}
& \, & \frac{dP}{dt}=-\beta P-qf(t), \\
& \, &  \dot{l}^2-2l\ddot{l}-2\beta l\dot{l}= \Omega_{R}^2 l^2\left[1+\frac{P^2}{M_{0}^2}\right]-\frac{1}{\alpha},
\label{cc-l}
\end{eqnarray}
where the momentum $P(t)=M_{0} l_{0} \dot{X}/l(t)$,
$\Omega_{R} = 1/(\sqrt{\alpha} l_{0})$ with $\alpha=\pi^2/12$ is the so-called Rice
frequency,
and 
$M_0=8$, $q=2\pi$  and $l_0=1$ are, respectively, the dimensionless 
kink mass, 
topological charge and unperturbed width.

Equation (\ref{cc-p}) can be solved exactly, and in the
large time limit ($t\gg \beta^{-1}$) yields
\begin{equation}\nonumber\label{4}
P(t)=-\sqrt{\epsilon}[a_{1}\sin(\delta t+\delta_{0}-\chi_{1})+a_{2}\sin(m\delta t+\delta_{0}+\theta-\chi_{2})],
\end{equation}
where $\epsilon$ is merely a rescaling parameter in
the perturbation expansion, to be determined later;
$\chi_{1}=\arctan\left({\delta}/{\beta}\right)$, $ \quad \chi_{2}=\arctan\left({m\delta}/{\beta}\right)$,
$a_{1}={q\epsilon_1}/{\sqrt{\epsilon(\beta^2+\delta^2)}},$ and 
$a_{2}={q\epsilon_2}/{\sqrt{\epsilon(\beta^2+m^2\delta^2)}}.$
The change of variable $g(t)^2=l(t)$ leads to 
an Ermakov-type equation for $g(t)$,
given by
\begin{equation}\label{6}
\ddot{g}+\beta\dot{g}+\left[\left(\frac{\Omega_{R}}{2}\right)^2+
\left(\frac{\Omega_{R}}{2M_{0}}\right)^2P^2(t)\right] g =
\frac{1}{4\alpha g^3},
\end{equation}

Let us recall that directed motion from ac driving was possible already for a 
single harmonic in the undamped sine-Gordon system \cite{Niurka1}, but was
suppressed by damping. Therefore, the 
interesting issue is to find net directed motion from ac driving 
in the presence of damping, i.e.,
$\beta\neq 0$. In that case, 
Eq.\ (\ref{cc-l}) cannot be solved\cite{Niurka2,Niurka3}, and therefore
we will study it by
a perturbative expansion, 
\begin{equation}
\label{7}
l(t)=l_{0}+\epsilon l_{1}(t)+\epsilon^2 l_{2}(t)+\cdots. 
\end{equation}
At order $O(\epsilon)$, we obtain from Eq.\ (\ref{6})
\begin{equation}\label{10}
\ddot{l}_{1}(t)+\beta \dot{l}_{1}(t)+\Omega_{R}^2 l_{1}(t)
=-{\Omega_{R}^2}P^{2}(t) l_{0}/{2\epsilon M_{0}^{2}}.
\end{equation}
If we now
substitute the expression of $P(t)$
into (\ref{10}), we find
\begin{eqnarray}\label{15}
\ddot{l}_{1}(t)+\beta \dot{l}_{1}(t)+\Omega_{R}^2 l_{1}(t)&=&
A_{1}+A_{2}\cos(2\delta t+2\delta_{0}-2\chi_{1})
+A_{3}\cos(2m\delta t+2\delta_{0}+2\theta-2\chi_{2})\\ \nonumber
&+&
A_{4}\cos[(m-1)\delta t +\theta -(\chi_{2}-\chi_{1})]-
A_{4}\cos[(m+1)\delta t+ 2\delta_{0}+ \theta -(\chi_{2}+\chi_{1})],
\end{eqnarray}
where
$A_{1}=-A_{2}-A_{3}$, and
$A_{2}={\Omega_{R}a_{1}^2}/{4\sqrt{\alpha }M_{0}^2},\>
A_{3}={\Omega_{R}a_{2}^2}/{4\sqrt{\alpha }M_{0}^2},\>
A_{4}=-{\Omega_{R}a_{1}a_{2}}/{2\sqrt{\alpha }M_{0}^2}.$
By inspection, we conclude from Eq.\ (\ref{15}) that
$l_{1}(t)$ contains harmonics of frequencies $2 \delta$,
$2 m \delta$ and $(m \pm 1) \delta$. This is a very important conclusion
and we will come back to it below. 

After transients elapse, we find
\begin{eqnarray} \label{16}
l_{1}(t)&=&\frac{A_{1}}{\Omega_{R}^2}+
\frac{A_{2}\sin(2\delta t+2\delta_{0}-2\chi_{1}+\tilde{\theta}_{2})}
{\sqrt{(\Omega_{R}^2-4\delta^2)^2+4\beta^2\delta^2}}+
\frac{A_{3}\sin(2m\delta t+2\delta_{0}+2\theta-2\chi_{2}+
\tilde{\theta}_{2m})}{\sqrt{(\Omega_{R}^2-4m^2\delta^2)^2+
4m^2\beta^2\delta^2}}\\ \nonumber
&&+\frac{A_{4}\sin[(m-1)\delta t +\theta -(\chi_{2}-\chi_{1})+
\tilde{\theta}_{m-1}]}{\sqrt{(\Omega_{R}^2-(m-1)^2\delta^2)^2+
\beta^2(m-1)^2\delta^2}}
-\frac{A_{4}\sin[(m+1)\delta t + 2\delta_{0}+
\theta -(\chi_{2}+\chi_{1})+\tilde{\theta}_{m+1}]}
{\sqrt{(\Omega_{R}^2-(m+1)^2\delta^2)^2+\beta^2(m+1)^2\delta^2}}, 
\end{eqnarray}
where
$\tilde{\theta}_{m}=
\arctan\left[\left({\Omega_{R}^2-m^2\delta^2}\right)/{m\beta\delta}\right].$
A cumbersome but otherwise 
trivial calculation yields the harmonics 
contained in $l_{2}(t)$,
collected in Table \ref{tab1}. 
\begin{table}
\caption{Harmonic content of the first contributions to the
perturbative expansion of $l(t)$.}\label{tab1}
\begin{center}
\begin{tabular}{|c|c|c|}
\hline
 harmonic & $l_{1}$ & $l_{2}$ \\
\hline
$m$ & $2\delta$, $2m\delta$, $(m\pm 1)\delta$ & $2\delta$, $4\delta$, $2m\delta$,
$4m\delta$, $(m\pm 1)\delta$, \\
\, & \, & $2(m\pm 1)\delta$, $(m\pm 3)\delta$, $(3m\pm 1)\delta$\\
\hline
2 & $\delta$, $2\delta$, $3\delta$, $4\delta$ & $\delta$, $2\delta$, $3\delta$, $4\delta$, $5\delta$, $6\delta$, $7\delta$, $8\delta$\\
\hline
3 & $2\delta$, $4\delta$, $6\delta$  &  $2\delta$, $4\delta$, $6\delta$, $8\delta$, $10\delta$,  $12\delta$ \\
\hline
4 & $2\delta$, $3\delta$, $5\delta$, $8\delta$ &  $\delta$,  $2\delta$, $3\delta$,  $4\delta$, $5\delta$, $7\delta$,\\
\, & \, & $9\delta$, $11\delta$, $13\delta$, $16\delta$\\
\hline
\end{tabular}
\end{center}
\end{table}

The next step is to obtain the velocity averaged over one period
of the driving. To this end, we use the definition of the 
momentum, namely 
\begin{equation}
\label{mom}
\Pi(t)=\int_{-\infty}^{\infty}dx\,\phi_t\phi_x,
\end{equation}
and find that, for our collective coordinate {\em Ansatz}, 
\begin{equation}\label{1}
\langle\dot{X}(t)\rangle=\frac{1}{T}\int_{0}^{T}\frac{P(t)l(t)}{M_{0}l_{0}}dt.
\end{equation}
Considering the expansion  (\ref{7}), this expression
can be written as
\begin{eqnarray}\label{13}
\nonumber
\langle\dot{X}(t)\rangle&=&\frac{1}{T}\int_{0}^{T}\frac{P(t)(l_{0}+
\epsilon l_{1}(t)+\epsilon^2 l_{2}(t)+...)}{M_{0}l_{0}}dt\\
&=&\langle\dot{X}_{0}(t)\rangle+\epsilon \langle\dot{X}_{1}(t)\rangle+\epsilon^2 \langle\dot{X}_{2}(t)\rangle+...
\end{eqnarray}
At $O(\epsilon^{0})$, the averages $\langle P(t)\rangle$
and $\langle\dot{X}_{0}(t)\rangle$ vanish trivially; therefore, 
net kink motion can only arise
in next order.  
By straightforward calculations from
Eqs.\ (\ref{16}) and (\ref{1}) we find for $m=2$ that, for large enough times,
\begin{eqnarray}
\epsilon\langle\dot{X}_{1}\rangle=
\frac{q^3\Omega_{R}^2\epsilon_{1}^2\epsilon_{2}}{8M_{0}^{3}
(\beta^2+\delta^2)\sqrt{\beta^2+4\delta^2}}\left(\frac{2\cos[\delta_{0}-
\theta+(\chi_{2}-2\chi_{1})-\tilde{\theta}_{1}]}{\sqrt{(\Omega_{R}^2-
\delta^2)^2+\beta^2\delta^2}} \right. \label{17}
\left.
-\frac{\cos[\delta_{0}-\theta+(\chi_{2}-2\chi_{1})+
\tilde{\theta}_{2}]}{\sqrt{(\Omega_{R}^2-4\delta^2)^2+
4\beta^2\delta^2}}\right).
\end{eqnarray}
It is now clear that, for Eq.\ (\ref{17}) to be valid, and our
perturbative expansion (\ref{7}) to hold, as $\epsilon$ has to be
small, the prefactor on the right-hand side of (\ref{17}) has 
to be much smaller than 1, which establishes a priori the range
of applicability of our theory (although generally speaking 
collective coordinate theories usually apply for larger perturbations
than predicted\cite{SIAM}).
This is the final result of the collective coordinate approach. 
We now proceed to discuss the predictions we can deduce from 
this expressions and to verify the results by comparing to 
numerical simulations. 

\section{Discussion}

To begin with, the result obtained in Eq.\ (\ref{17}) predicts that for $m=2$
kinks should travel with a non-zero velocity, in agreement with the reports in
Refs.\ \citenum{Flach,SZ}. If we fix $\delta_0=\pi/2$, the case studied by those
authors, we find a sinusoidal dependence on the relative phase $\theta$ exactly
as reported in those previous works. At some specific values of this relative 
phase, the velocity vanishes, meaning that parameters can be chosen as to move
in any desired direction or to keep the kink oscillating around some point. 

Secondly, the expression (\ref{17}) has more implications. Indeed, it can 
be seen that the velocity has an additional sinusoidal dependence on the 
initial phase, $\delta_0$. This is a new result, not discussed by Flach 
{\em et al.} and Salerno and Zolotaryuk, that shows that relative and initial
phase play exchangeable roles. That can be immediately seen by a change of 
variables in the original sine-Gordon equation (\ref{sG}): taking $\delta_0$
to be $\delta t_0$, we find that changing the time variable in the fashion
$t'=t+t_0$ leads to the conclusion that an initial phase $\delta_0$ is 
equivalent to a relative phase $\theta'=\theta-(m-1)\delta_0$ for 
a kink displaced from $x_0$ to $x_0+Vt_0$. 

\begin{figure}
\begin{center}
\includegraphics[height=7cm]{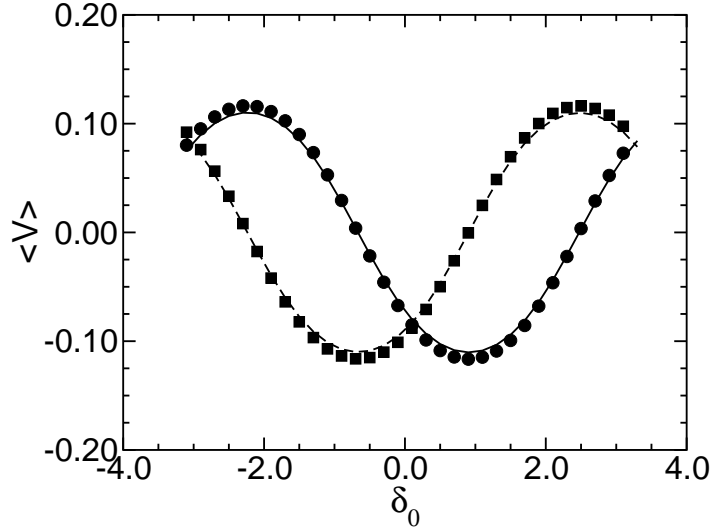}
\caption{\label{fig:faseinicial}
Dependence of the kink velocity on the initial phase. Parameters are
$\epsilon_1=\epsilon_2=0.2$, $\beta=0.05$, $\delta=0.1$. Relative phase $\theta=\pi/2$:
solid line, CC theory; filled circles, simulation results.
Relative phase $\theta=0$:
dashed line, CC theory; squares, simulation results.
}
\end{center}
\end{figure}

Before proceeding any further, it is important to check the validity of the
two conclusions we have just presented. After all, our collective coordinate
theory is basically a variational
approach, and in addition we have also treated perturbatively the resulting
equations. Therefore, numerical simulations of the full sine-Gordon equation
(\ref{sG}) are needed to verify the predictions of the calculations in the
previous section. To this end, we carried out such simulations (see 
Ref.\ \citenum{prl} for details on the numerical procedure) and the 
outcome is plotted in Fig.\ \ref{fig:faseinicial}. 
We see that the agreement between theory and simulation is perfect, and there 
indeed is the double dependence on both the initial and the relative phase as
found in Eq.\ (\ref{17}). This fully confirms our approach and even our 
predictions in a quantitative manner. 

At this point, it is interesting to consider the following issue, which arises
if we want to consider our model as a ratchet: Looking at the sinusoidal dependence
on the initial phase, it may be concluded that the system is not a ratchet 
because an average over any possible initial phase would in principle
lead to zero mean velocity. However, this is not the case: Recall that,
as mentioned in the previous paragraph, changing the initial phase implies 
changing the relative phase as well. Then we can see, by inspecting Eq.\ (\ref{17}),
that the kink velocity depends on the {\em difference $\delta_0-\theta$}, which 
is the same for all choices of $\delta_0$. Hence, the kink velocity is the 
same for all $\delta_0$, and averaging does not suppress the ratchet effect. 

With the $m=2$ results of our approach well established by the 
comparison with the simulations, let us look at the $m=3$ case.
Once again, our theoretical results agree with and explain the 
numerical findings in Refs.\ \citenum{Flach,SZ}: the prediction
from Eq.\ (\ref{17}) is that the velocity is zero and therefore
that there is no net directed motion of the kink, as observed
in those previous papers. To be sure, we have observed that if 
the driving force is strong enough (not for any driving value, as
in the ratchet effect) directed motion is observed (as has 
been reported in \cite{SZ}), but the system
is so largely distorted that one cannot conclude that the kink is
moving as a coherent entity (it looks as though the motion is 
driven by the kink {\em wings}), and therefore that is an altogether
different question. 

\begin{figure}
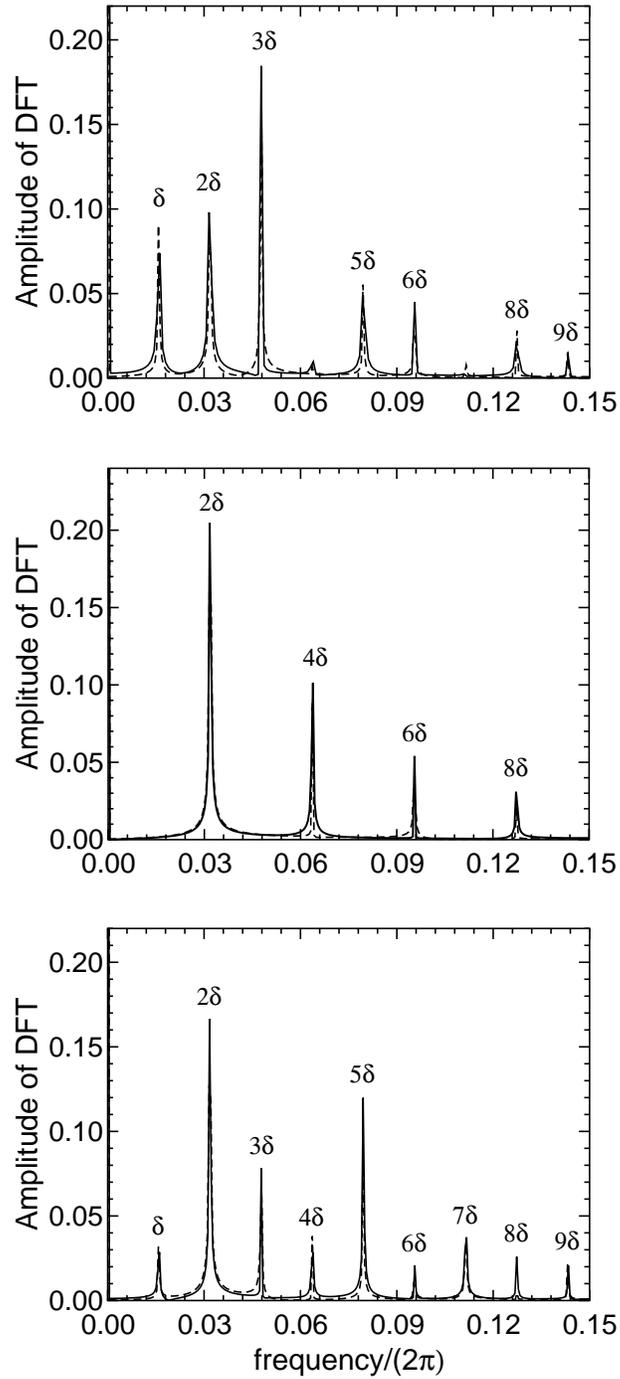

\begin{center}
\includegraphics[height=6cm]{dftm2.eps}\\[1mm]
\includegraphics[height=6cm]{dftm3.eps}\\[1mm]
\includegraphics[height=6cm]{dftm4.eps}
\caption{\label{fig:dft}
Discrete Fourier Transform of the kink width.
Upper panel: $m=2$; middle panel: $m=3$;
lower panel: $m=4$.
Solid line: amplitude measured in simulations. Dashed line:
numerical integration of the CC equations.
Parameters are as in Fig.\ \ref{fig:faseinicial}
for relative phase $\theta=\pi/2$ and $\delta_0=-2.5$.}
\end{center}
\end{figure}

Why is net directed motion suppressed for $m=3$? The explanation of
this rule is transparent from Table \ref{tab1} and constitutes the
most relevant success of our approach. For $m=3$, the 
frequencies of the ac force (or the momentum) are odd 
harmonics of $\delta$, whereas the width of the kink 
(governed by the internal mode) oscillates only with even 
harmonics. Figure \ref{fig:dft} shows clearly and with very 
high accuracy that the simulations behave exactly as predicted by 
the collective coordinate equations. Note that in Fig.\
\ref{fig:dft} we plot results obtained for $m=4$ as well, 
a situation in which we recover the ratchet effect and the 
kink moves towards a preferred direction for any value of the
ac driving. 

This remark directs us to both the mechanism 
responsible for the appearance of directed motion and the 
statement of a selection rule based on it. {\em The mechanism is 
the coupling between the translation of the kink and the 
internal mode}. This is reflected in 
Eqs.\ (\ref{cc-p}) and (\ref{cc-l}), which show that
the ac driving $f(t)$ acts on the kink width through $P^2(t)$, whereas $P(t)$
itself is in turn
inversely proportional to $l(t)$. This coupling is the responsible
for the net kink motion, but for it to be actually possible,
{\em the harmonic content of the effective
force $P^2(t)$ acting on the width
degree of freedom must be able to resonate with it.}
This is evident from Eq.\ (\ref{1}), in which the integral is nonzero
only if $l(t)$ contains at least one of the harmonics of $P(t)$.
It is important to realize that this condition is much more restrictive
than that found in Ref.\ \citenum{Willis}, where only the necessity of $l(t)$
being a dynamic variable was pointed out, but no conditions on its 
frequency content or any other feature were obtained. We have just seen that this
is indeed necessary, but that additional, crucial resonance conditions
have to be fulfilled.

\section{Effects of noise}

\begin{figure}
\begin{center}
\includegraphics[height=7cm]{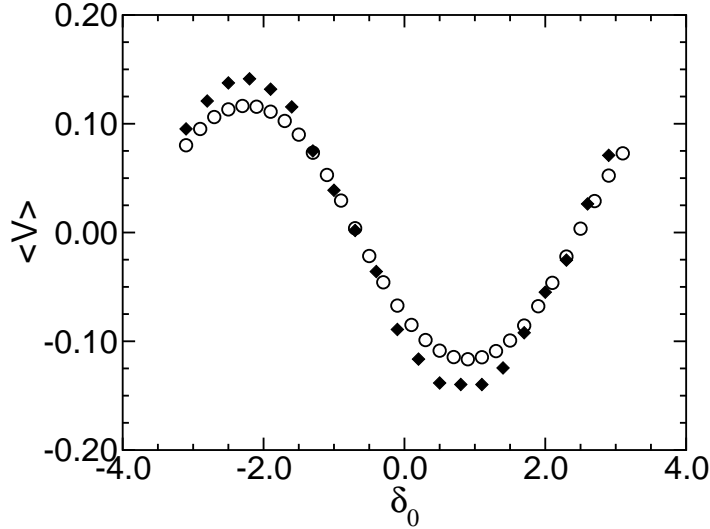}
\caption{\label{fig:ruido}
Dependence of the kink velocity on the initial phase for
relative phase $\theta=\pi/2$ in the deterministic ($D=0$,
empty circles) and the stochastic ($D=0.03$, diamonds) cases.
Other parameters are as in Fig.\ \ref{fig:faseinicial}.}
\end{center}
\end{figure}

As a final step to set our results on firm grounds, we studied 
the robustness of the kink directed motion against noisy perturbations
(much in the same way as in Ref.\ \citenum{SZ}).
To this end, we considered the Langevin equation 
\begin{equation}
\phi_{tt} - \phi_{xx} + \sin(\phi) = -\beta \phi_{t} + f(t) + \xi(x,t),
\label{ssG}
\end{equation}
where $\xi(x,t)$ is a gaussian white noise of zero mean and 
correlations $\langle\xi(x,t)\xi(x',t')\rangle=D\delta(x-x')\delta(t-t')$.
In this case, the equation was studied only by numerical simulations, which 
were carried out using a Heun scheme \cite{revruido}. 
While one could in principle think that this
noise would suppress the initial phase dependence, Fig.\ \ref{fig:ruido} shows
that the opposite is the case: the {\em noise enhances the dependence on the
initial phase}, increasing the maximum values of the velocity while keeping
the same general sinusoidal dependence and the location of the zeros. It
is tempting to conclude from this plot that the noise, at least if it is
not very large, assists the process of energy transfer between the width
and the translation degrees of freedom, activating it.

\begin{figure}
\begin{center}
\includegraphics[height=10cm,angle=270]{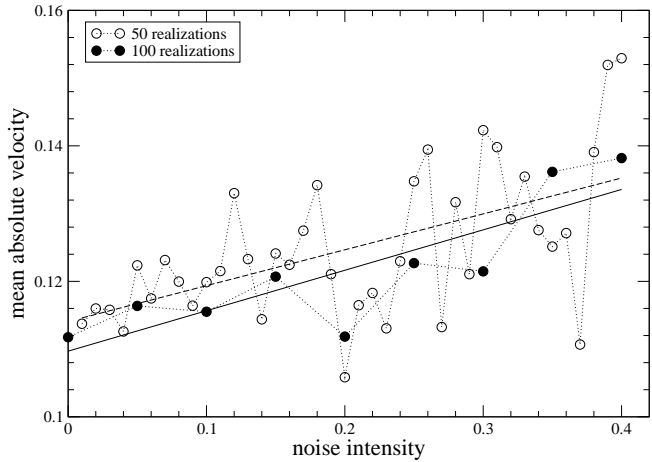}
\caption{\label{fig:sr}
Mean absolute value of the kink center velocity vs noise intensity. Parameters
are 
$\epsilon_1=\epsilon_2=0.2$, $\beta=0.05$, $\delta=0.1$,
initial phase $\delta_0=-0.31831$, and relative phase $\theta=-\pi/2$. 
Realizations in the averages are as indicated. The straight lines are
linear fits of slopes 0.053 (dashed, for the 50 realizations set) and
0.060 (solid, for the 100 realizations set).}
\end{center}
\end{figure}

Having verified, as intended, that the ratchet effect is robust to the 
effects of noise, we find it interesting to look any further on this 
activation phenomenon in its own right. Therefore, we looked at one 
particular point in Fig.\ (\ref{fig:ruido}) where the velocity was 
larger (in absolute value) and performed numerical simulations 
varying only the noise intensity, leaving all other parameters unchanged. 
The results are summarized in Fig.\ \ref{fig:sr}. 
{}The first conclusion we can draw from this plot is that the noise has 
indeed a nontrivial activation effect on the kink motion.
As can be seen, 
the velocity exhibits an increasing trend with 
the noise intensity, although the dependence is rather noisy. To verify 
the trend we have made two linear fits to the two sets of simulation 
data, finding comparable slope values of 0.053 and 0.060. 
Upon increasing the noise further, we found that kink-antikink pairs 
were spontaneously generated in the system, which eventually collide with
the original one and render the simulation useless (from the viewpoint 
of measuring the kink velocity). However, 
it is clear that if we 
were able to study even larger values of the noise, the velocity has to decrease 
as diffusion becomes the dominant process, effectively suppressing 
the ac driving effects. In that case, asymptotically, the 
velocity must go to zero as noise increases, and we 
would be observing a manifestation of a phenomenon 
of much interest during the last decade, namely {\em stochastic resonance}. 
Stochastic resonance was (very much like ratchets) first found in point-like
systems \cite{SR}, but it has been generalized to extended systems in the last few 
years (see, e.g., Ref.\ \citenum{SSR}). In this respect, it is interesting
to recall Ref.\ 
\citenum{Katja}, where thermal resonances as a function of temperature were 
observed, to our knowledge for the first time. By comparing the system in 
that previous work with the present one, we believe that we can indeed 
expect the appearance of resonant phenomena. 
We thus plan to carry out further simulations with different parameters, 
in particular, with larger damping, in order to suppress the nucleation
of kink-antikink pairs and be able to observe the dynamics of a single 
kink for larger noises. Analytical calculations 
will also be desirable to ascertain the nature of
this activation process. 
In any event, what we have established is that 
net directed motion can be observed even for moderate and large noise 
intensities, which shows that the mechanism proposed here is extraordinarily 
robust. Work along these lines is in progress.

\section{Conclusions}

Regarding the deterministic part, the main conclusion of this work is the 
understanding of the condition for the appearance of net directed
motion in an ac driven, damped sine-Gordon system. Our collective coordinate
theory, thoroughly confirmed by the numerical simulations of the full 
partial differential equation, has allowed us to show that the motion 
arises from the coupling of the internal mode (effectively arising from 
odd phonon contributions) to the translation of the 
kink. As a byproduct, we have found the physical reasons for the symmetry 
requirements first proposed in Refs.\ \citenum{Flach,SZ} for the existence
of this ratchet-like phenomenon. Indeed, to make net motion possible, the 
indirect forcing arising from the internal mode influence has to resonate
with the available frequencies for the width of the kink itself. 
Interestingly, while this phenomenon should be expected in any other
soliton-bearing model with solitons possessing an internal mode 
(e.g., the $\phi^4$ model), for the sine-Gordon equation it is an 
apparent puzzle due to its lack of internal modes \cite{Niurka3}. 
However, as explained above, (odd) phonons have been shown to give
rise to width oscillations very similar to those induced by an internal 
mode \cite{Niurka5}. We believe that what we are seeing is precisely the 
result of the action of those phonons, summarized in our approach through
the width variable $l(t)$. This conclusion reinforces, in turn, the interpretation
of Quintero and Kevrekidis \cite{Niurka5}, which will probably be useful in 
other instances of sine-Gordon problems where kink oscillations play a 
role. Our analytical calculations have also clarified the influence of the 
relative phase and the initial phase, establishing their equivalence. 
In this respect, we should stress the fact that the dependence of the 
kink velocity on the initial phase does not suppress the ratchet effect,
due precisely to this relationship with the relative phase and the 
result that the velocity is unchanged when changing the initial phase. 

With respect to the effects of noise, the conclusions are not that 
definite. We have certainly established the robustness of the ratchet
effect under the influence of noise, even for rather large noise 
values. In addition, we have verified that for moderate noises, the
velocity of the directed motion increases with noise intensity, up to 
the point in which thermally generated kink-antikink pairs do not allow
us to follow a single kink. The existence of this activation process 
along with the fact that diffusion must dominated at larger noises
have led us to conjecture that the system could exhibit some form
of stochastic resonance.
In order to confirm this interpretation, much more numerical
and analytical work is needed. Apparently, the motion takes place through
the same mechanism, so the two collective coordinate equations are 
likely to provide a good starting point to understand the phenomenon. 
However, it is also known that the diffusion of sine-Gordon solitons is
controlled by a diffusion constant which depends both linearly and 
quadratically on the noise intensity in the full partial differential 
equation \cite{xxxx}. This is another factor that could be playing a
role in the observed nonmonotonous dependence of the velocity on the 
noise. It is then clear that the problem is still far from understood 
and that we cannot offer a final conclusion about it at this stage. 

From a more applied point of view, this work 
may have important consequences for applications
as a way of separating, e.g., fluxons in long Josephson junctions
\cite{ext-app}. Interestingly, such superconducting devices (and related
ones, see e.g. Ref.\ \citenum{Hanggi_a} and references therein) provide the
best possible laboratory to verify our results.
This experimental confirmation is crucial in order
to ascertain their applicability. Given the accuracy with which the sG
equation describes long Josephson junctions, and the fact that an external
force like the one proposed in this and earlier works \cite{Flach,SZ} is
easy to implement, we hope that the corresponding measurements will soon
be carried out. As a bonus, the possible existence of stochastic resonance
would then be easier to explore experimentally. 
Finally, the applicability of this internal mode mechanism 
may have other, far-reaching implications in very different kinds of 
systems. Indeed, the original motivation of this work is the modelling of
molecular motors such as kinesin. It is well known now that kinesin 
``walks" along microtubules by simultaneously changing its shape 
(see Ref.\ \citenum{kinesin} and references therein) to allow one ``head''
to overcome the other. The internal mode mechanism proposed here is 
clearly reminiscent of the (obviously much more complicated) way kinesin
moves. As another example, we want to mention the recent experiments on 
kink-induced transport in granular media by Moon {\em et al.}\cite{prl2},
in which it has been observed directed motion of kinks separating 
convection rolls when the system is under the influence of two harmonic 
drivings similar to those considered here. While the direct comparison of
our simple model with these phenomena is out of the question, we do 
believe that the paradigm of coupling to internal modes could help
understand the experiments from a more mesoscopic viewpoint. 

\acknowledgments     
 
AS has been
supported by the Ministerio de Ciencia y Tecnolog\'\i a of Spain
through grant
BFM2003-07749-C05-01.
NRQ has been
supported by the Ministerio de Ciencia y Tecnolog\'\i a of Spain
through grant
BFM2001-3878-C02, 
by the Junta de Andaluc\'{\i}a under 
project FQM-0207, by DAAD (Germany)
A0231253/Ref.\ 314, and, along with LMM and FGM,
by the International Research Training Group
`Nonequilibrium Phenomena and Phase Transitions in Complex Systems'
(DFG, Germany).
KL has been supported by the Engineering Research Program of
the Office of Basic Energy Sciences at the U.\ S.\ Department of Energy
under Grant No.\ DE-FG03-86ER13606.



\begin{thebibliography}{88}
\bibitem{Scott} A.\ C.\ Scott, {\em Nonlinear Science} (Oxford University,
Oxford, 1999).
\bibitem{Reimann} P.\ Reimann, Phys.\ Rep.\ \textbf{361}, 57 (2002).
\bibitem{Hanggi_a} R. Dean Astumian and P.\ H\"anggi, Phys.\ Today {\bf 55}
(11), 33 (2002).
\bibitem{Prost} F.\ J\"ulicher, A.\ Adjari and J.\ Prost, Rev.\ Mod.\
Phys.\ {\bf 69}, 1269 (1997).
\bibitem{Linke} H.\ Linke, ed., {\em Ratchets and Brownian Motors: 
Basics, Experiments and Applications}, Appl.\ Phys.\ A \textbf{75} 
(2002), special issue.
\bibitem{Feynman} R.\ P.\ Feynman, R.\ B.\ Leighton, M.\ Sands, {\em 
The Feynman Lectures in Physics}, Vol.\ I, ch.\ 46 (Addison-Wesley,
Reading, 1963).
\bibitem{Magnasco} M.\ O.\ Magnasco, Phys.\ Rev.\ Lett.\ {\bf 71}, 1477 
(1993).
\bibitem{Prost2} J.\ Prost, J.-F. Chauwin, L.\ Peliti and A.\ Adjari, 
Phys.\ Rev.\ Lett.\ {\bf 72}, 2652 (1994).
\bibitem{0d} 
P.\ Jung, J.\ G.\ Kissner, and P.\ H\"anggi, Phys.\ 
Rev.\ Lett.\ \textbf{76}, 3436 (1996);
T.\ E.\ Dialynas, K.\ Lindenberg, and G.\ P.\ Tsironis, 
Phys.\ Rev.\ E \textbf{56}, 3976 (1997);
J.\ L.\ Mateos, Phys.\ Rev.\ Lett.\ \textbf{84}, 258 
(2000).
\bibitem{SIAM} A.\ S\'anchez and A.\ R.\ Bishop, SIAM Rev.\ \textbf{40}, 579
(1998).
\bibitem{Marchesoni} F.\ Marchesoni, Phys.\ Rev.\ Lett.\ \textbf{77},
2364 (1996). 
\bibitem{Tsironis} A. V. Savin, G. P. Tsironis and A. V. Zolotaryuk, 
Phys.\ Lett.\ A \textbf{229}, 279 (1997); Phys.\ Rev.\ E \textbf{56}, 
2457 (1997).
\bibitem{ext-th} P.\ Reimann, R.\ Kawai, C.\ Van den Broeck, and P.\ H\"anggi,
Europhys.\ Lett.\ \textbf{45}, 545 (1999);
C.\ Van den Broeck, I.\ Bena, P.\ Reimann, and J.\ Lehmann,
Ann.\ Phys.\ (Leipzig) \textbf{9}, 713 (2000);
J.\ Buceta, J.\ M.\ Parrondo, C.\ Van den Broeck, and F.\ J.\ de la Rubia,
Phys.\ Rev.\ E \textbf{61}, 6287 (2000);
G.\ Costantini and F.\ Marchesoni, 
Phys.\ Rev.\ Lett.\ \textbf{87}, 114102 (2001);
Z.\ Zheng, M.\ C.\ Cross, and G.\ Hu, Phys.\ Rev.\ 
Lett.\ \textbf{89}, 154102 (2002);
G.\ Costantini, F.\ Marchesoni, and M.\ Borromeo,
Phys.\ Rev.\ E \textbf{65}, 051103 (2002);
L.\ M.\ Flor\'\i a, F.\ Falo, P.\ J.\ Mart\'\i nez, and J.\ J.\ Mazo,
Europhys.\ Lett.\ \textbf{60}, 174 (2002).
\bibitem{salquin} M.\ Salerno and N.\ R.\ Quintero, Phys.\ Rev.\ E
\textbf{65}, 025602 (2002).
\bibitem{ext-app} G.\ Carapella and G.\ Costabile, Phys.\ Rev.\ Lett.\ 
\textbf{87}, 077002 (2001);
E.\ Goldobin, A.\ Sterck, and D.\ Koelle, Phys.\ Rev.\ 
E \textbf{63}, 031111 (2001);
F.\ Falo, P.\ J.\ Mart\'\i nez, J.\ J.\ Mazo, T.\ P.\ Orlando, K.\ Segall, E.\ Tr\'\i as,
in Ref.~\citenum{Linke}, p.\ 263.
\bibitem{Flach2} S.\ Flach, O.\ Yevtushenko, and Y.\ Zolotaryuk, 
Phys.\ Rev.\ Lett.\ \textbf{84}, 2358 (2000).
\bibitem{Hanggi_b} I.\ Goychuk and P.\ H\"anggi, in {\em Stochastic Processes in
Physics, Chemistry and Biology}, edited by J.\ A.\ Freund and T.\ P\"oschel 
(Lecture Notes in Physics, no.\ 557, Springer-Verlag, Berlin, 2000).
\bibitem{Hanggi_c} I.\ Goychuk and P.\ H\"anggi, J.\ Phys.\ Chem.\ B {\bf 105},
6642 (2001).
\bibitem{Flach} S.\ Flach, Y.\ Zolotaryuk, A.\ E.\ Misoshnichenko, and M.\ V.\ Fistul, 
Phys.\ Rev.\ Lett.\  \textbf{88}, 184101 (2002). 
\bibitem{SZ} M.\ Salerno and Y.\ Zolotaryuk, Phys.\ Rev.\ E \textbf{65}, 
056603 (2002). 
\bibitem{Niurka1} N.\ R.\ Quintero and A.\ S\'anchez, Phys.\ Lett.\ A 
\textbf{247}, 161 (1998); Eur.\ Phys.\ J.\ B \textbf{6}, 133 (1998). 
\bibitem{Niurka2} N.\ R.\ Quintero, A.\ S\'anchez, and F.\ G.\ Mertens, 
Phys.\ Rev.\ Lett.\  \textbf{84}, 871 (2000). 
\bibitem{Niurka4} N.\ R.\ Quintero, A.\ S\'anchez, and F.\ G.\ Mertens, 
Phys.\ Rev.\ E \textbf{62}, 5695 (2000).
\bibitem{Willis} C.\ R.\ Willis and M.\ Farzaneh, {\tt arXiv:cond-mat/0212125}.
\bibitem{Niurka3} N.\ R.\ Quintero, A.\ S\'anchez, and F.\ G.\ Mertens, 
Phys.\ Rev.\ E \textbf{62}, R60 (2000). 
\bibitem{Niurka5} N.\ R.\ Quintero and P.\ G.\ Kevrekidis, Phys.\ Rev.\ 
E \textbf{64}, 056608 (2001).
\bibitem{Cilla} S.\ Cilla, F.\ Falo, and L.\ M.\ Flor\'\i a, Phys.\
Rev.\ E \textbf{63}, 031110 (2001).
\bibitem{prl} L.\ Morales-Molina, N.\ R.\ Quintero, F.\ G.\ Mertens,
and A.\ S\'anchez, Phys.\ Rev.\ Lett.\ {\bf 91}, 234102 (2003).
\bibitem{Rice} M.\ Salerno and A.\ C.\ Scott, Phys.\ Rev.\ B {\bf 26}, 
2474 (1982); M.\ J.\ Rice, Phys.\ Rev.\ B \textbf{28}, 3587 (1983).
\bibitem{revruido} M.\ San Miguel and R.\ Toral, in 
{\em Nonequilibrium Structures VI}, edited by 
E.\ Tirapegui (Kluwer, Dordrecht, 1998).
\bibitem{SR} L.\ Gammaitoni, P.\ H\"anggi, P.\ Jung, 
and F.\ Marchesoni, Rev.\ Mod.\ Phys.\ {\bf 70}, 223 (1998).
\bibitem{SSR} J.\ Garc\'\i a-Ojalvo and J.\ M.\ Sancho, 
{\em Noise in spatially extended systems} 
(Springer-Verlag, New York, 1999).
\bibitem{Katja} R.\ Reigada, A.\ Sarmiento, and
  K.\ Lindenberg, Phys.\ Rev.\ E {\bf 63}, 66113 (2001).
\bibitem{xxxx} N.\ R.\ Quintero, A.\ S\'anchez and F.\ G.\ Mertens
Eur.\ Phys.\ J.\ B {\bf 16}, 361 (2000).
\bibitem{kinesin} A.\ Yildiz, M.\ Tomishige, R.\ D.\ Vale and P.\ R.\ Selvin, 
Science {\bf 303}, 676 (2004).
\bibitem{prl2} S.\ J.\ Moon, D.\ I.\ Goldman, J.\ B.\ Swift,
and H.\ L.\ Swinney, Phys.\ Rev.\ Lett.\ {\bf 91}, 134301 (2003).
\end{thebibliography}
\end{document}